\documentclass[12pt]{article}
\newtheorem{lemma}{Lemma}
\newtheorem{proposition}{Proposition}
\newtheorem{theorem}{Theorem}
\newtheorem{corollary}{Corollary}

\def\Tr{\mathop{\rm Tr}\nolimits}

\def\dif{{\rm d}}
\def\ci{{\rm{i}}}

\catcode`\@=11 \long\def\@makefntext#1{ \protect\noindent \hbox to
3.2pt {\hskip-.9pt
$^{{\ninerm\@thefnmark}}$\hfil}#1\hfill}        

 \def\@makefnmark{\hbox to 0pt{$^{\@thefnmark}$\hss}}  

\def\ps@myheadings{\let\@mkboth\@gobbletwo
\def\@oddhead{\hbox{}
\rightmark\hfil\ninerm\thepage}
\def\@oddfoot{}\def\@evenhead{\ninerm\thepage\hfil
\leftmark\hbox{}}\def\@evenfoot{}
\def\sectionmark##1{}\def\subsectionmark##1{}}

\renewenvironment{thebibliography}[1]
    {\begin{list}{$^{\arabic{enumi}}$}
    {\usecounter{enumi}\setlength{\parsep}{0pt}
\setlength{\leftmargin 1.25cm}{\rightmargin 0pt}
     \setlength{\itemsep}{0pt} \settowidth
    {\labelwidth}{#1.}\sloppy}}{\end{list}}

\newcounter{itemlistc}
\newcounter{romanlistc}
\newcounter{alphlistc}
\newcounter{arabiclistc}

\def\@citex[#1]#2{\if@filesw\immediate\write\@auxout
    {\string\citation{#2}}\fi
\def\@citea{}\@cite{\@for\@citeb:=#2\do
    {\@citea\def\@citea{,}\@ifundefined
    {b@\@citeb}{{\bf ?}\@warning
    {Citation `\@citeb' on page \thepage \space undefined}}
    {\csname b@\@citeb\endcsname}}}{#1}}

\newif\if@cghi
\def\cite{\@cghitrue\@ifnextchar [{\@tempswatrue
    \@citex}{\@tempswafalse\@citex[]}}
\def\citelow{\@cghifalse\@ifnextchar [{\@tempswatrue
    \@citex}{\@tempswafalse\@citex[]}}
\def\@cite#1#2{{$\null^{#1}$\if@tempswa\typeout
    {IJCGA warning: optional citation argument
    ignored: `#2'} \fi}}

\def\fnt#1#2{\footnotetext{\kern-.3em
    {$^{\mbox{\sevenrm #1}}$}{#2}}}

 1 
scaled\magstep 1  1
 
scaled\magstephalf 
  
 \font\ninerm=cmr9

\begin{document}
\title{On the invariant symmetries of the $\mathcal{D}$-metrics}
\author{Joan Josep Ferrando$^1$ and Juan Antonio S\'aez$^2$}
\date{\empty}

\maketitle \vspace*{-0.5cm}
\begin{abstract}
We analyze the symmetries and other invariant qualities of the
$\mathcal{D}$-metrics (type D aligned Einstein Maxwell solutions
with cosmological constant whose Debever null principal directions
determine shear-free geodesic null congruences). We recover some
properties and deduce new ones about their isometry group and about
their quadratic first integrals of the geodesic equation, and we
analyze when these invariant symmetries characterize the family of
metrics. We show that the subfamily of the Kerr-NUT solutions are
those admitting a Papapetrou field aligned with the Weyl tensor.
\end{abstract}

\vspace*{2mm}

\begin{center}
PACS: 04.20.Cv, 04.20.-q
\end{center}

\vspace*{2cm} \noindent $^1$ Departament d'Astronomia i
Astrof\'{\i}sica, Universitat de Val\`encia, E-46100 Burjassot,
Val\`encia, Spain.
E-mail: {\tt joan.ferrando@uv.es}\\
$^2$ Departament de Matem\`atiques per a l'Economia i l'Empresa,
Universitat de Val\`encia, E-46022 Val\`encia, Spain. E-mail: {\tt
juan.a.saez@uv.es}
\newpage

\section{Introduction}

Explicit integration of the Einstein-Maxwell equations with
cosmological constant has been achieved by several authors for the
family of aligned type D metrics whose null principal directions
define shear-free geodesic
congruences\cite{ple-dem,weir-kerr,deb-kam-mc} (see also references
therein). This family of solutions has been named the ${\cal
D}$-metrics.\cite{deb-kam-mc} They can be deduced from the
Pleba\'nsky and Demia\'nski\cite{ple-dem} line element by means of
several limiting procedures (see the Stephani {\em et al.}
book\cite{kramer} and references therein, and the recent paper by
Griffiths and Podolsk\'y\cite{griff-pod} for a detailed analysis).

The family of the ${\cal D}$-metrics contains significant and
well-known solutions of the Einstein equations, like the
Reissner-Nordstr\"om and the Kerr-Newman black holes and their
vacuum limit, the Schwarzschild and Kerr solutions, as well as,
other well-known space-times that generalize them, such as the
charged Kerr-NUT solutions. The generalized C-metrics, that describe
two accelerated black holes moving in opposite
directions,\cite{kin-walker} also belong to this family.

Elsewhere\cite{fs-EM-align} we have given the Rainich theory for
type D aligned Einstein-Maxwell solutions with (or without)
cosmological constant. The interesting nature of this study is that
offers an intrinsic and explicit characterization of the ${\cal
D}$-metrics by means of algebraic conditions on the curvature
tensor. Now, in the present work we analyze another important
quality of the ${\cal D}$-metrics: they admit several {\em invariant
symmetries}.

Although the existence of these symmetries can be shown starting
from the plain expression of the metric line element, their
invariant character and their close relation with the curvature
tensor become more evident if one presents them without explicit
integration of the Einstein-Maxwell equations. We can quote some
historical references that afford results in this sense.

The existence of a 2+2 conformal Killing tensor in type D vacuum
metrics and in their charged counterpart was shown by Walker and
Penrose.\cite{wape} They also proved that a full Killing tensor
exists in the charged Kerr black hole. Starting from this result
Hougston and Sommers\cite{hs1} studied the general conditions under
which this Killing tensor may be constructed.

In a subsequent work, Hougston and Sommers\cite{hs2} proved that the
symmetries of these space-times arise from properties of the
curvature tensor. They gave the expression of a complex Killing
vector as a (differential) concomitant of an aligned conformal
Killing-Yano bivector, and they showed that this Killing vector
degenerates (it defines a unique real Killing vector) if, and only
if, the metric admits a Killing tensor.\cite{hs2} Moreover, in this
degenerate case, at least one other Killing vector exists that
commutes with the first one.

At this point a paper by Collinson and Smith\cite{cs} showed that,
under the assumptions of the papers by Hughston  and Sommers, if a
Killing tensor exits, then the space-time also admits a Killing-Yano
2-form. A similar result was also obtained by Stephani.\cite{ste}

In this work we revisit all these invariant symmetries of the ${\cal
D}$-metrics and we analyze how these geometric properties
characterize this family of space-times. We also point out that the
inheriting property of the (source) electromagnetic field is related
to the commutativity of the 2-dimensional isometry group that the
${\cal D}$-metrics admit in accord with a result by Michalski and
Wainwright.\cite{mich-wain} Moreover, we show that this feature can
be stated as a condition on the curvature tensor.

Another interesting result of this work concerns the Papapetrou
fields defined by the Killing vectors. If $\xi$ is a Killing vector,
the {\it Killing 2--form} $\nabla \xi$ is closed and, in the vacuum
case, it is a solution of the source-free Maxwell equations. Because
this fact was pointed out by Papapetrou,\cite{pap} the covariant
derivative $\nabla \xi$ has also been called the {\it Papapetrou
field}.\cite{faso1}

Metrics admitting an isometry were studied by considering the
properties of the associated Killing 2--form,\cite{deb1,deb2} and
this approach was extended to the spacetimes with an homothetic
motion.\cite{mc1,mc2} The vacuum solutions with an isometry have
been classified by considering the algebraic structure of the
Killing 2--form,\cite{faso1,faso2} and several extensions have been
developed for homothetic and conformal motions\cite{stee,lud} and
for non vacuum solutions.\cite{faso3}

In the Kerr geometry the principal directions of the Killing 2--form
associated with the timelike Killing vector coincide with the two
double principal null (Debever) directions of the Weyl
tensor.\cite{faso1} This means that the Killing 2--form is a Weyl
principal bivector. This fact has been remarked upon by
Mars\cite{mars} who has also shown that it characterizes the Kerr
solution under asymptotic flatness behavior.

A question naturally arises: can all the vacuum solutions with this
property of the Kerr metric be determined?
Elsewhere\cite{fsI1a,fsI1b} we have shown that the Petrov-Bel type I
vacuum space-times with this characteristic admit a 3--dimensional
group of isometries of Bianchi type I or II and that a close
relationship exists between the Weyl principal directions and the
isometry group.

In the present work we fulfil a similar study for Petrov-Bel type D
metrics. Thus, we show here that the Kerr-NUT solutions are the type
D vacuum metrics with a non null Killing 2--form aligned with the
Weyl geometry. Actually this property also holds for the charged
Kerr-NUT solutions with cosmological constant.

The paper is organized as follows. In Sec. II we introduce the
notation and some useful concepts about 2+2 space-time structures
and non null Maxwell fields that we need in the work. The non
conformally flat metrics admitting a conformal Killing-Yano 2-form
are analyzed and characterized in Sec. III. In Sec. IV we show that
if the metric is invariant under the divergence of this conformal
Killing-Yano tensor, then both the Ricci and Weyl tensor have the
same algebraic type as the ${\cal D}$-metrics. In Sec. V we study
several equivalent conditions, related to the invariance of the
electromagnetic field, that satisfy the ${\cal D}$-metrics and we
show that these properties also hold for a slightly wider family of
space-times. In Sec. VI we offer two characterizations of the ${\cal
D}$-metrics, one of them based on the properties of the conformal
Killing-Yano tensor. The other one is intrinsic and explicit and
imposes conditions on the curvature tensor. Finally, Sec. VII is
devoted to analyzing the Kerr-NUT space-times in detail and to
studying the equivalence of several geometric properties that
characterize them, like that of admitting an aligned Papapetrou
field.

\section{Some notation and useful concepts}
\label{sec-2}

Let $(V_4,g)$ be an oriented space-time of signature $\{ -, +,+,+
\}$. The metric $G$ on the space of 2--forms is $G=\frac{1}{2} g
\wedge g$, $\wedge$ denoting the double-forms exterior product, $(A
\wedge B)_{\alpha \beta \mu \nu} = A_{\alpha \mu} B_{\beta \nu} +
A_{\beta \nu} B_{\alpha \mu} - A_{\alpha \nu} B_{\beta \mu} -
A_{\beta \mu} B_{\alpha \nu}$. If $F$ and $H$ are 2--forms, $(F,H)$
denotes the product with $G$:
$$
(F,H) \equiv G(F,H) = \frac14 G_{\alpha \beta \lambda \mu} F^{\alpha
\beta} H^{\lambda \mu} = \frac12 F_{\alpha \beta} H^{\alpha \beta}\,
.
$$

If $A$ and $B$ are two 2--tensors, we denote $A \cdot B$ the tensor
with components $(A \cdot B)_{\alpha \beta} = A_{\alpha}^{\ \mu}
B_{\mu \beta}$. Moreover, $[A,B]$ and $\{A,B\}$ denote the
commutator and anti-commutator, respectively:
$$
[A,B] = A \cdot B - B \cdot A  \, ,\qquad \{A,B\} = A \cdot B + B
\cdot A \, .
$$

A self--dual 2--form is a complex 2--form ${\cal F}$ such that
$*{\cal F}= \textrm{i}{\cal F}$, where $*$ is the Hodge dual
operator. We can associate biunivocally with every real 2--form $F$
the self-dual 2--form ${\cal F}=\frac{1}{\sqrt{2}}(F-\textrm{i}*F)$.
For short, we here refer to a self--dual 2--form as a {\it SD
bivector}. The endowed metric on the 3-dimensional complex space of
the SD bivectors is ${\cal G}=\frac{1}{2}(G-\textrm{i} \; \eta)$,
$\eta$ being the metric volume element of the space-time.

If $F$ is a 2--form and $P$ and $Q$ are double-2--forms, $\Tr P$,
$P(F)$ and $P \circ Q$ denote, respectively, the scalar, the 2--form
and the double-2--form given by:
$$
\Tr P \equiv  \frac12 P_{\alpha \beta}^{\ \ \alpha \beta} \, ,
\qquad P(F)_{\alpha \beta} \equiv \frac12 P_{\alpha \beta}^{\ \ \mu
\nu} F_{\mu \nu} \, , \qquad (P \circ Q)_{\alpha \beta \rho \sigma}
\equiv \frac12 P_{\alpha \beta}^{\ \ \mu \nu} Q_{\mu \nu \rho
\sigma} \, .
$$
Moreover, we write $P^2 = P \circ P$.

Every double-2--form, and in particular the Weyl tensor $W$, can be
considered as an endomorphism on the space of the 2--forms. The
restriction of the Weyl tensor on the SD bivectors space is the {\em
self-dual Weyl tensor} and it is given by:
$$
{\cal W} \equiv {\cal G} \circ W \circ {\cal G} = \frac12(W - \ci
*W) \, .
$$

\subsection{Space-time 2+2 almost-product structures}

On the space-time, a 2+2 almost-product structure is defined by a
time-like field of planes $V$ and its space-like orthogonal
complement $H$. Let $v$ and $h= g-v$ be the respective projectors
and let $\Pi = v-h$ be the {\it structure tensor}. A 2+2 space-time
structure is also determined by the {\it canonical} unitary 2-form
$U$, volume element of the time-like plane $V$. Then, the respective
projectors are $v=U^2$ and $h = -(*U)^2$, where $U^2 = U \cdot U$.

In working with 2+2 structures it is useful to introduce the
canonical SD bivector ${\cal U} \equiv \frac{1}{\sqrt{2}} (U - {\rm
{i}} *U )$ associated with $U$, that satisfies $2{\cal U}^2 = g$
and, consequently, $({\cal U},{\cal U}) = -1$. In terms of ${\cal
U}$, the structure tensor is $ \Pi = 2{\cal U} \cdot \bar{{\cal
U}}$, where $\bar{\ }$ denotes complex conjugate.

The 2+2 almost-product structures can be classified by taking into
account the invariant decomposition of the covariant derivative of
the structure tensor $\Pi$ or, equivalently, according to the
foliation, minimal or umbilical character of each
plane.\cite{cf,fsD,cfsK} We will say that a structure is integrable
when both, $V$ and $H$ are a foliation. We will say that the
structure is minimal (umbilical) if both, $V$ and $H$ are minimal
(umbilical).

All these geometric properties of a 2+2 structure admit a
kinematical interpretation,\cite{cf,fsD} and they can be stated in
terms of some first order differential concomitants of
$U$.\cite{fsD,cfsK} Now we summarize some of these results which we
will use in the following sections. If $i({\xi})$ denotes the
interior product with a vector field $\xi$, and $\delta$ the
exterior codifferential, $\delta = *d*$, we have the following
lemma:\cite{fsD}

\begin{lemma}  \label{lem-umb}
Let us consider the 2+2 structure defined by the canonical
2--form $U$. The three following conditions are equivalent: \\
(i) The structure is umbilical\\
(ii) The canonical SD bivector ${\cal U}=\frac{1}{\sqrt{2}} (U -
{\rm i} *U)$ satisfies:
\begin{equation}
\Sigma[U] \equiv \nabla {\cal U} - i({\xi}) {\cal U} \otimes {\cal
U} - i({\xi}){\cal G}=0 \, , \qquad \xi \equiv \delta {\cal U} \, .
\label{umb}
\end{equation}
(iii) The principal directions of $U$ determine shear-free geodesic
null congruences.
\end{lemma}
The umbilical condition (\ref{umb}) can be written, equivalently, in
terms of the structure tensor $ \Pi = 2{\cal U} \cdot \bar{{\cal
U}}$:\cite{fsY}
\begin{equation}
\sigma[\Pi] \equiv {\bf s}\{2\nabla\Pi + \Pi(\nabla \cdot \Pi)
\otimes \Pi - (\nabla \cdot \Pi) \otimes g \} = 0 \, , \label{sigma}
\end{equation}
where ${\bf s}\{t\}$ stands for the total symmetrization of a tensor
$t$, and $(\nabla \cdot \Pi)_{\alpha} = \nabla_{\lambda}
\Pi^{\lambda}_{\, \alpha}$.

On the other hand, a 2+2 structure is minimal (respectively,
integrable) if, and only if, the {\em expansion vector} $\Phi$
(respectively, the {\em rotation vector} $\Psi$) vanishes.\cite{fsD}
These concomitants of $U$ are given by:
\begin{equation} \label{Phi-Psi}
\begin{array}{l}
\Phi  \equiv \Phi[U] \equiv i(\delta U) U - i(\delta *U)*U \, , \\[1mm]
\Psi \equiv \Psi[U] \equiv - i(\delta U) *U - i(\delta *U)U \, .
\end{array}
\end{equation}
In terms of the structure tensor $\Pi$, the expansion and rotation
vectors are, respectively:\cite{fsY}
\begin{equation} \label{PhiPsiPi}
\begin{array}{l}
\Phi = \Phi[\Pi] \equiv - \frac{1}{2} \Pi(\nabla \cdot \Pi) \, , \\
\Psi = \Psi[\Pi] \equiv  \frac{3}{2} *(\nabla \Pi \cdot \Pi) \, ,
\end{array}
\end{equation}
where, for a 2-tensor $A$ and a vector $x$, $A(x)_{\mu} =  A_{\mu
\nu} x^{\nu}$, and we put $*t$ to indicate the action of the Hodge
dual operator on the skew-symmetric part of a tensor $t$.

Finally, we show a property on the integrable directions in a
2-plane that we will use in this work. It is known\cite{fsD,cfsK}
that the 2-plane $H$ is integrable (that is, $V$ is a foliation) if,
and only if, $i(\delta U)*U = 0$. This fact ensures that every
direction in $H$ is integrable. Now, let us suppose that $H$ is not
integrable but that $H$ contains one integrable direction $X$, that
is,
\begin{equation}  \label{integrable-direction}
i(X) U = 0 \, , \qquad \dif X \wedge X = 0 \, .
\end{equation}
From here we obtain $(U, \dif X) =0$, and then
$$
(X, \delta U) = (U, \dif X) - \delta i(X)U = 0 \, .
$$
Then, $X$ and $h(\delta U)$ are orthogonal directions in the 2-plane
$H$ or, equivalently, $X$ and $i(\delta U)\!*\!U$ are collinear,
that is, $\,  X\wedge i(\delta U)\!*\!U = 0$. A similar result can
be obtained by exchanging $H$ and $V$. Thus, we can state:
\begin{lemma} \label{lemma-integ}
Let $U$ be the canonical 2-form of a 2+2 structure $(V,H)$. Then:

(i) The 2-plane $V$ (respectively, $H$) is integrable if, and only
if, $i(\delta \!*\!U)U=0$ (respectively, $i(\delta U)\!*\!U=0$).

(ii) If $V$ (respectively, $H$) is not integrable and an integrable
direction exists in $V$ (respectively, $H$), then this direction is
given by $i(\delta \!*\!U)U$ (respectively, $i(\delta U)\!*\!U$).

(iii) If $X$ is an integrable direction in $V$ (respectively, $H$),
then
$$
(X,\, \delta \!*\!U) = 0 \, , \qquad  ({\rm respectively}, \ (X,\,
\delta U) = 0) \, .
$$
\end{lemma}

\subsection{Non null Maxwell fields}

A non-null 2-form $F$ takes the canonical expression $\, F =
e^{\phi}[\cos \psi \, U \! + \sin \psi *\!U]$, where $U$ is a simple
and unitary 2--form that we name {\it geometry} of $F$, $\phi$ is
the {\it energetic index} and $\psi$ is the {\it Rainich index}. The
intrinsic geometry $U$ determines a 2+2 almost-product structure
defined by the {\em principal planes}, the time-like one $V$ whose
volume element is $U$, and its space-like orthogonal complement $H$.

The energy ({\it Maxwell-Minkowski}) tensor $T$ associated with an
electromagnetic field $F$ is minus the traceless part of its square
and, in the non-null case, it depends on the intrinsic variables
$(U, \phi)$:
\begin{equation}
T \equiv -\frac{1}{2}[F^2+*F^2] = -\frac{1}{2}e^{2\phi}[U^2+*U^2] =
-\kappa \Pi  \, .  \label{TF}
\end{equation}

The symmetric tensor (\ref{TF}) has the principal planes of the
electromagnetic field as eigen-planes and their associated
eigen-values are $\pm \kappa$, with $2 \kappa = \sqrt{\Tr T^2} =
e^{2\phi}$.

In terms of the {\em intrinsic elements} $(U,\phi,\psi)$ of a
non-null Maxwell field, the source-free Maxwell equations, $\delta F
=0$, $\delta *F =0$, take the expression:\cite{rai,cff}
\begin{equation} \label{maxwell-rainich}
\dif \phi = \Phi[U] \, , \qquad \qquad   \dif \psi = \Psi[U]  \, .
\end{equation}

When $F$ is solution of the source-free Maxwell equations, one says
that $U$ defines a {\it Maxwellian structure}. Besides, when the
Maxwell-Minkowski energy tensor $T$ associated with a non-null
2--form is divergence--free, the underlying 2+2 structure is said to
be {\em pre-Maxwellian}.\cite{deb} The conservation of $T$ is
equivalent to the first of the Maxwell-Rainich equations
(\ref{maxwell-rainich}).\cite{fsY} Then, from these equations we
obtain the following result:\cite{rai,fsY}
\begin{lemma}
(i) A {\em 2+2} structure is Maxwellian if, and only if, the
expansion and the rotation are closed 1--forms, namely the canonical
2--form $U$ satisfies:
\begin{equation}\label{max2}
\mbox{\rm d} \Phi[U]  = 0 \, , \qquad  \qquad \mbox{\rm d} \Psi[U] =
0 \, .
\end{equation}

(ii) A {\em 2+2} structure is pre-Maxwellian if, and only if, the
canonical 2--form $U$ satisfies the first equation in {\rm
(\ref{max2})}.
\end{lemma}

We can collect the expansion vector $\Phi$ and the rotation vector
$\Psi$ in a complex vector $\chi$ with a simple expression in terms
of the canonical SD bivector ${\cal U}$. Indeed, we have:
\begin{equation} \label{chi}
\Phi + \ci \Psi = 2 \chi \, , \qquad \chi \equiv \chi[{\cal U}] =
i({\xi}) {\cal U} \, , \quad \xi \equiv \delta{\cal U}  \, .
\end{equation}
Then, conditions (\ref{max2}) that characterize a Maxwellian
structure can be written as $\dif \chi = 0$.

\section{Metrics admitting a non null conformal Killing-Yano 2-form $A$}
\label{sec-CKY-A}

The ${\cal D}$-metrics are Petrov-Bel type D solutions whose
principal null directions define shear-free geodesic congruences.
Moreover, the source of the gravitational field is a non null
Maxwell field whose principal directions are those of the Weyl
tensor (aligned Einstein-Maxwell solution). Thus, using the
terminology introduced in Sec. 2, the Weyl principal 2+2 structure
is umbilical and Maxwellian. These geometric restrictions also hold
for the vacuum limit as well as for type D metrics with vanishing
Cotton tensor.\cite{fsD} Moreover, elsewhere\cite{fsY} we have shown
that these properties characterize the non conformally flat
space-times admitting a conformal Killing-Yano 2-form. In this
section we summarize what this means and we make the conditions it
imposes on the curvature tensor explicit.

A Conformal Killing-Yano (CKY) 2-form is a solution $A$ to the
conformal invariant extension to the Killing-Yano equation. The
conformal Killing-Yano equation takes the form:\cite{tach}
\begin{equation} \label{CKYd}
\nabla_{(\alpha} A_{\beta) \mu} =  g_{\alpha \beta} a_{\mu} -
a_{(\alpha} g_{\beta) \mu }  \, ,
\end{equation}
where the 1--form $a$ is given by the codifferential of $A$: $3a = -
\delta A$.

If $A$ is a CKY 2-form, the scalar $\nu = A(n,p)$ is constant along
an affinely parameterized null geodesic with tangent vector $n$,
where $p$ is a vector orthogonal to  the geodesic and satisfying $n
\wedge \nabla_n p =0$. In particular, we can take $p = A(n)$, which
satisfies these restrictions as a consequence of the CKY equation.
Then, the scalar $A^2(n,n)$ is a quadratic first integral of the
null geodesic equation, so that, the traceless part of $A^2$ is a
second rank conformal Killing tensor, that is, a 2+2 symmetric
tensor $P$ solution to the conformal Killing equation
\begin{equation} \label{CKd}
\nabla_{(\alpha} P_{\beta \mu)} = g_{(\alpha \beta} b_{\mu)} \, ,
\end{equation}
where the 1-form $b$ is defined by: $3b = \nabla \cdot P$.

Elsewhere\cite{cfsK} we have shown (see also a related previous
result by Hauser and Malhiot)\cite{h-m-2} that a metric $g$ which
admits a 2+2 conformal Killing tensor $P$ is conformal to a metric
admitting a totally geodesic structure. This means that an umbilical
and pre-Maxwellian structure exists in this
space-time.\cite{fsY,cfsK} Moreover, the full Maxwellian character
of this structure is equivalent to the conformal Killing tensor
being the traceless square of a CKY 2-form.\cite{fsY}

On the other hand, we have studied\cite{fsU} the restrictions that
the existence of a Maxwellian and umbilical (two shear-free geodesic
null congruences) structure imposes on the curvature tensor. In what
follows, we will go on to use some of these results which we now
summarize in three lemmas:\cite{fsY,cfsK,fsU}
\begin{lemma} \label{lemma-CKY-1}
A non conformally flat space-time admits a non null conformal
Killing-Yano 2-form $A$ if, and only if, the Weyl tensor is
Petrov-Bel type D and the Weyl principal structure is umbilical and
Maxwellian, that is to say, the Weyl principal bivector $U$
satisfies:
\begin{equation} \label{CKY2}
\Sigma[U]=0 \, ;    \qquad  \qquad \mbox{\rm d} \Phi[U]  = 0 \, ,
\qquad  \qquad \mbox{\rm d} \Psi[U]  = 0 \, ,
\end{equation}
where $\Sigma[U]$, $\Phi[U]$ and $\Psi[U]$ are given in {\em
(\ref{umb})} and {\em (\ref{Phi-Psi})}.
\end{lemma}

\begin{lemma} \label{lemma-CKY-2}
If a non conformally flat space-time admits a non null conformal
Killing-Yano 2-form $A$, and $U$ is the principal bivector of the
(type D) Weyl tensor, then two functions $(\phi,\psi)$ exist such
that $\mbox{d} \phi = \Phi[U]$,
$\mbox{d} \psi = \Psi[U]$. Moreover:\\
(i) The $2$-form $F = e^{\phi}[\cos \psi U + \sin \psi *U]$ is a
(test) Maxwell field aligned with the Weyl tensor and whose
principal directions define shear-free geodesic congruences.\\
(ii) In terms of the electromagnetic variables $(U, \phi, \psi)$,
the intrinsic variables of the CKY tensor are $(U, -\phi/2,
-\psi/2)$, that is, it takes the expression $A = e^{-(\phi/2)}[\cos
(\psi/2) U - \sin (\psi/2) \! * \!U]$
\end{lemma}

\begin{lemma} \label{lemma-CKY-3}
In a non conformally flat space-time which admits a non null
conformal Killing-Yano tensor $A$ the Ricci tensor $R$ satisfies
${\cal S} = [R, {\cal U}]$, ${\cal U}$ being the principal bivector
of the (type D) Weyl tensor and where:
\begin{equation} \label{2S}
{\cal S} = {\cal S}[{\cal U}] \equiv {\cal L}_{\xi} g - \xi \otimes
\chi - \chi \otimes \xi\, , \qquad \xi = \delta {\cal U}, \quad \chi
= i({\xi}){\cal U} \, .
\end{equation}
\end{lemma}

This lemma shows a close relationship between the Ricci tensor $R$
and the Weyl principal bivector ${\cal U}$. Nevertheless, this
limitation of the Ricci tensor does not necessarily restrict its
algebraic type. Thus, the presence of a non null CKY tensor
restricts the Weyl tensor to be of Petrov-Bel type D (or O), but the
Ricci tensor may be algebraically general.

\section{Metrics invariant under the complex vector $\delta {\cal A}$}
\label{sec-KV-deltaA}

The CKY equation (\ref{CKYd}) is invariant under the Hodge duality,
so that, if $A$ is a CKY tensor, $*A$ is a CKY tensor too.
Consequently, the SD bivector ${\cal A} = \frac12 (A - \ci *A)$
satisfies the CKY equation. In terms of the electromagnetic
variables $(U, \phi, \psi)$ of the Maxwell field given in lemma
\ref{lemma-CKY-2}, the SD bivector associated with the CKY 2-form
$A$ takes the expression:
\begin{equation} \label{sd-CKY}
{\cal A} = e^{-\frac12(\phi + \ci \psi)} {\cal U} \, , \qquad \quad
{\cal U} \equiv \frac{1}{\sqrt{2}} (U - \ci *U) \, .
\end{equation}

Hougston and Sommers\cite{hs2} showed that, for the ${\cal
D}$-metric family of solutions, the divergence of the (self-dual)
CKY bivector ${\cal A}$ is a complex Killing vector field. Now we
analyze the generic metrics where this property holds and we show
that it imposes strong restrictions on the Ricci tensor.

Let us consider the complex vector ${\cal Z} \equiv \delta {\cal
A}$. From (\ref{sd-CKY}), and considering the $\xi$ and $\chi$ given
in (\ref{chi}), a straightforward calculation leads to:
\begin{equation}  \label{ckv}
{\cal Z} \equiv \delta {\cal A} = \frac32 \, e^{-\frac12(\phi + \ci
\psi)} \xi = 3 \, e^{-\frac12(\phi + \ci \psi)} i({\chi}) {\cal U}
\, .
\end{equation}
From this relation and expression (\ref{2S}) of ${\cal S}$, and
taking into account that $\dif \phi + \ci \dif \psi = 2 \chi$, we
obtain that ${\cal L}_ {\cal Z} g = 0$ if, and only if,  ${\cal S}$
vanishes identically. But lemma \ref{lemma-CKY-3} implies that this
condition states that the Ricci tensor commutes with ${\cal U}$:
$[{\cal U},R]=0$. Finally, an algebraic calculation shows that this
requisite restricts the Ricci tensor to be of algebraic type
[(11)(11)], and the two eigen-planes are the principal planes of the
bivector ${\cal U}$. Thus, we obtain:
\begin{proposition} \label{prop-KV}
The non conformally flat space-times admitting a conformal
Killing-Yano bivector ${\cal A}$ whose divergence either vanishes or
is a (complex) Killing vector are those with the
following properties:\\
(i) The Weyl tensor is Petrov-Bel type D and the null principal
directions define shear-free geodesic congruences.\\
(ii) The Ricci tensor is of type {\em [(11)(11)]} and aligned with
the Weyl tensor, that is,
\begin{equation} \label{ricci-pi-lambda}
R = - \kappa \Pi + \Lambda g  \, ,
\end{equation}
where $\Pi = 2 {\cal U} \cdot \bar{{\cal U}}$, ${\cal U}$
being the Weyl principal bivector.\\
(iii) There exists a (test) Maxwell field aligned with the Weyl and
Ricci tensors, $F = e^{\phi}[\cos \psi U + \sin \psi * \!U]$, whose
intrinsic variables $(U, \phi, \psi)$ determine the CKY bivector as
expression (\ref{sd-CKY}).
\end{proposition}

Note that the space-times characterized in the proposition above
have a Ricci tensor that is, from an algebraic point of view, of
(aligned) electromagnetic type $- \kappa \Pi$ plus a term
proportional to the metric tensor, $\Lambda g$. But the invariant
scalar $\Tr R = 4 \Lambda$ is not, necessarily, constant and then
the energy momentum tensor of the Maxwell field $F$ is not the
source of the Einstein equations, $2 \kappa \not= e^{2\phi}$. Thus,
the family of the ${\cal D}$-metrics is a strict subset of this
family of space-times.

\section{Space-times with the CKY ${\cal A}$ invariant under the
Killing vector $\delta {\cal A}$} \label{sec-CKY-invariant}

If an Einstein-Maxwell solution admits a Killing vector $Z$ then, as
a consequence of the field equations, the electromagnetic energy
tensor $T$ is also invariant under $Z$. Nevertheless, the
electromagnetic field $F$ does not necessarily inherit the symmetry.
More precisely, $F$ satisfies:\cite{ray-t,mich-wain,coll}
\begin{equation}
{\cal L}_Z F = k *\!F \, ,
\end{equation}
where $k$ is a constant when $F$ is a non null
field.\cite{ray-t,mich-wain} When $F$ is a null field, $k$ is either
a constant (if $F$ is non-integrable) or a function that determines
the front waves (if $F$ is integrable).\cite{coll}

Michalski and Wainwright\cite{mich-wain} showed that if the
Einstein-Maxwell space-time admits a 2-parameter orthogonally
transitive Abelian group of isometries, then the electromagnetic
field inherits the symmetries. It is known that such an Abelian
group exists in the case of the ${\cal D}$-metrics and,
consequently, the electromagnetic field which is the source of these
solutions is invariant under the Killing vectors.

Here we analyze the inheriting property from another viewpoint.  In
order to gain a better understanding of this behavior of the ${\cal
D}$-metrics, let us consider the wider family of space-times
characterized in proposition \ref{prop-KV}. In this case only the
canonical geometry $U$ is, a priory, invariant under the Killing
vector ${\cal Z}$ because, as a consequence of the alignment, $U$ is
a Weyl concomitant (the principal bivector). Consequently, the
(test) electromagnetic field $F$ will be invariant under ${\cal Z}$
if, and only if, both, the energetic and Rainich indexes $(\phi,
\psi)$ are.

Besides, Maxwell equations (\ref{maxwell-rainich}) imply that the
energetic index $\phi$ (respectively, the Rainich index $\psi$) is
invariant under ${\cal Z}$ if, and only if, the expansion vector
$\Phi$ (respectively, rotation vector $\Psi$) is orthogonal to
${\cal Z}$, $(\Phi, {\cal Z})=0$ (respectively, $(\Psi, {\cal
Z})=0$).

On the other hand, from the expression (\ref{ckv}) of the Killing
vector ${\cal Z}$ and from the Maxwell equations
(\ref{maxwell-rainich}), and taking into account that the metric
concomitant $\xi = \delta {\cal U}$ is invariant under the Killing
vector ${\cal Z}$, one obtains:
\begin{equation} \label{lie-Z-Z}
[{\cal Z}, \bar{{\cal Z}}] = - ({\cal Z}, \bar{\chi}) \bar{{\cal Z}}
\, .
\end{equation}
If $Z_1$ and $Z_2$ denote the real and imaginary parts of ${\cal
Z}$, ${\cal Z} = Z_1 + \ci Z_2$, we have $[{\cal Z}, \bar{{\cal Z}}]
= - 2 \ci [Z_1, Z_2]$. Consequently, from (\ref{lie-Z-Z}), ${\cal
Z}$ determines a commutative 2-dimensional Killing algebra (or, a
unique Killing direction) if, and only if, $({\cal Z}, \bar{\chi}) =
0$. As a consequence of expression (\ref{ckv}) this orthogonality
condition states, equivalently, that ${\cal U} (\chi, \bar{\chi}) =
0$ . Taking the real and imaginary parts of this condition we get
$U(\Phi, \Psi) =
*U(\Phi, \Psi) = 0$. Moreover, it is evident that ${\cal U} (\chi,
\Phi) = 0$ iff ${\cal U} (\Phi, \Psi) = 0$ iff ${\cal U} (\chi,
\Psi) = 0$, that is, the expansion vector $\Phi$ is orthogonal to
${\cal Z}$, $(\Phi, {\cal Z})=0$, iff the rotation vector $\Psi$ is,
$(\Psi, {\cal Z})=0$.

Finally both, the (test) Maxwell field $F$ and the CKY tensor $A$,
have the same intrinsic elements $(U, \phi, \psi)$ and,
consequently, $F$ is invariant under the Killing vector ${\cal Z}$
iff $A$ is invariant. All these considerations allow us to state:

\begin{proposition} \label{prop-A-Z}
For the non conformally flat space-times admitting a conformal
Killing-Yano bivector ${\cal A}$ whose divergence ${\cal Z} \equiv
\delta {\cal A} \not= 0$ is a
(complex) Killing vector, the following conditions are equivalent:\\
(i) The CKY bivector ${\cal A} = e^{-\frac12(\phi + \ci \psi)} {\cal
U}$ is invariant under the Killing vector ${\cal Z}$.\\
(ii) The (test) Maxwell field $F = e^{\phi}[\cos \psi U + \sin \psi
*U]$ is invariant under the Killing vector ${\cal Z}$.\\
(iii) The energetic index $\phi$ is invariant under the Killing
vector ${\cal Z}$.\\
(iv) The Rainich index $\psi$ is invariant under the Killing vector
${\cal Z}$.\\
(v) The expansion vector $\Phi$ is orthogonal to the Killing
vector ${\cal Z}$, $(\Phi, {\cal Z}) = 0$.\\
(vi) The rotation vector $\Psi$ is orthogonal to the Killing
vector ${\cal Z}$, $(\Psi, {\cal Z}) = 0$.\\
(vii) $[Z_1, Z_2]=0$, where ${\cal Z} = Z_1 + \ci Z_2$, that is,
${\cal Z}$ determines either a commutative 2-dimensional Killing
algebra or a unique Killing direction.\\
(viii) $U(\Phi, \Psi) = *U(\Phi, \Psi) = 0$, that is, the projection
of the expansion and rotation vectors on the principal planes are
collinear.
\end{proposition}

Let us remark the different sort of (equivalent) points in
proposition \ref{prop-A-Z}. The first one exclusively involves the
CKY bivector ${\cal A}$: it is invariant under its divergence
$\delta {\cal A}$.

The second one is related to the inheriting property of the
(aligned) electromagnetic field and, in agreement with the Michalski
and Wainwright result,\cite{mich-wain} it is a consequence of the
commutative character of the Killing algebra (point (vii)).
Nevertheless, in our reasoning here, $F$ is a {\em test}
electromagnetic field and not, necessarily, the source of the
Einstein-Maxwell equations (the energetic index $\phi$ is not, a
priory, a metric concomitant). Moreover, the inheriting property
holds even when the Killing vector determines a unique Killing
direction, that is, the existence of an Abelian group is not
generically imposed although, actually, a 2-dimensional commutative
algebra also exists in this degenerate case (see section
\ref{sec-Kerr-NUT}).

Conditions (iii) and (iv) show that the invariance of a sole
electromagnetic invariant scalar implies the invariance of the other
one, and then of the full electromagnetic field. This fact clarifies
the inheriting property for the ${\cal D}$-metrics, where the
energetic index $\phi$ is, a priory, a metric concomitant and,
consequently, it is invariant under the Killing vectors.

Maxwell equations allow us to express the above invariant properties
(iii) and (iv) (differential conditions) as algebraic restrictions
on the expansion and rotation vectors (conditions (v) and (vi)).
Finally, condition (vii) exclusively involves the Weyl principal
bivector $U$ and, thus, it is an intrinsic restriction on the
space-time.

\section{Characterizing the ${\cal D}$-metrics}

In the previous sections we have studied some properties that
satisfy the non null CKY 2-form that the ${\cal D}$-metrics admit.
Now we analyze whether these restrictions are also sufficient
conditions, that is, if they characterize this family of
space-times.

Lemmas \ref{lemma-CKY-1} and \ref{lemma-CKY-2} (Sec. III) state that
the existence of the CKY 2-form in a non conformally flat space-time
restrict the Weyl tensor to be Petrov-Bel type D with the two
principal null directions defining shear-free geodesic congruences.

In proposition \ref{prop-KV} (Sec. IV) we have studied the condition
on the Ricci tensor to be of electromagnetic type
(\ref{ricci-pi-lambda}) and aligned with the Weyl tensor and we have
shown that it can be stated as a condition on the CKY  bivector
${\cal A}$ given in (\ref{sd-CKY}): its divergence is a complex
Killing vector.

On the other hand, for the Ricci tensor (\ref{ricci-pi-lambda}) the
contracted Bianchi identities take the form:
\begin{equation}
2 \kappa   \  \Phi = \dif \kappa  - \Pi(\dif \Lambda) \, .
\end{equation}
Consequently, from the first Maxwell-Rainich equation
(\ref{maxwell-rainich}), $- \kappa \Pi$ is the energy momentum
tensor of the Maxwell field $F = e^{\phi}[\cos \psi U + \sin \psi
*U]$ ($2 \kappa = e^{2\phi}$) if, and only if, $\Lambda$ is a
(cosmological) constant. Thus, up to this simple condition on the
Ricci tensor, the ${\cal D}$-metrics may be characterized with
qualities of a CKY tensor. More precisely, we have:
\begin{theorem} \label{teorema-CKY-D-metrics}
The ${\cal D}$-metrics are the non conformally flat space-times with
constant scalar curvature, $\dif \Tr R = 0$, admitting a non null
CKY bivector ${\cal A}$ whose divergence, $\delta {\cal A}$, either
vanishes or is a complex Killing vector.
\end{theorem}

It is worth remarking that the sole condition $\dif \Tr R = 0$
implies that $\phi$ is a metric concomitant and, consequently, it is
invariant under the Killing vector $\delta {\cal A}$. Then, taking
proposition \ref{prop-A-Z} into account, we can state:
\begin{corollary} \label{corol-D-invariant}
The ${\cal D}$-metrics satisfy each of the (equivalent) conditions
in proposition \ref{prop-A-Z}. In particular, the source
electromagnetic field $F$ is invariant under the Killing vector
$\delta {\cal A}$.
\end{corollary}

The characterization of the ${\cal D}$-metrics given in theorem
\ref{teorema-CKY-D-metrics} is not intrinsic because it imposes
conditions on a bivector ${\cal A}$ and, a priori, we do not know it
in terms of the metric tensor. Elsewhere\cite{fs-EM-align} we have
acquired a whole intrinsic and explicit characterization of the
${\cal D}$-metrics that involves both Ricci and Weyl concomitants,
and that has the suitable property of being purely algebraic in
these tensors.

A whole intrinsic and explicit characterization of a metric or a
family of metrics is quite interesting from a conceptual point of
view - and from a practical one - because it can be tested by direct
substitution of the metric tensor in arbitrary coordinates. Thus,
this approach is an alternative to the usual approach to the metric
equivalence problem. This and other advantages have been pointed out
elsewhere\cite{fsS} where this kind of identification has been
obtained for the Schwarzschild space-time as well as for all the
other type D static vacuum solutions. A similar study has been
carried out for a family of Einstein-Maxwell solutions that include
the Reissner-Nordstr\"{o}m metric.\cite{fsD}

In order to obtain intrinsic and explicit characterizations, as well
as having an intrinsic labeling of the metrics, we need  to express
these intrinsic conditions in terms of explicit concomitants of the
metric tensor. When doing this, the role played by the results on
the covariant determination of the eigenvalues and eigenspaces of
the Ricci tensor\cite{bcm} and the principal 2--forms and principal
directions of the Weyl tensor\cite{fsI,fms} is essential.

Now, in this section, we present another intrinsic and explicit
characterization of the ${\cal D}$-metrics, which is alternative to
that quoted above.\cite{fs-EM-align} The present one is differential
in the curvature tensor but it has an advantage: either it involves
the Weyl principal bivector ${\cal U}$ and, therefore, it can be
stated by means of conditions on the Weyl tensor, or it restricts
the principal structure tensor $\Pi$ and, therefore, it can be
stated by means of conditions on the sole Ricci tensor.

Let us consider first the characterization of the ${\cal D}$-metrics
by means of the principal bivector ${\cal U}$. Conditions
(\ref{CKY2}) in lemma \ref{lemma-CKY-1} assure the existence of a
CKY 2-form. On the other hand, if the tensor ${\cal S}$ given in
(\ref{2S}) vanishes, then the Ricci tensor takes the expression
(\ref{ricci-pi-lambda}), and proposition \ref{prop-KV} implies that
the the divergence of the CKY bivector is a Killing vector.
Therefore, from theorem \ref{teorema-CKY-D-metrics} we obtain:
\begin{proposition} \label{prop-char-U}
The ${\cal D}$-metrics are the non conformally flat space-times with
constant scalar curvature, $\dif \Tr R = 0$, that admit a unitary
bivector ${\cal U}$ satisfying:
\begin{equation} \label{charac-U}
\Sigma[{\cal U}]=0 \, ,    \qquad  \quad \mbox{\rm d} \chi[{\cal U}]
= 0 \, ,  \qquad \quad {\cal S}[{\cal U}] = 0 \, ,
\end{equation}
where $\Sigma[{\cal U}]$, $\chi[{\cal U}]$ and ${\cal S}[{\cal U}]$
are given, respectively, in {\em (\ref{umb})}, {\em (\ref{chi})}
and {\em (\ref{2S})}.\\
Moreover, the space-time is Pertov-Bel type D and ${\cal U}$ is the
principal bivector of the Weyl tensor.
\end{proposition}

From this proposition, and considering the results given
elsewhere\cite{fms} about the invariant characterization of the
Petrov-Bel type D metrics and the covariant determination of their
Weyl principal bivector, we obtain:

\begin{theorem} \label{Teor-charac-weyl}
The ${\cal D}$-metrics are the space-times with constant scalar
curvature, $\dif \Tr R = 0$, whose Weyl tensor satisfies the
algebraic conditions:
\begin{equation}  \label{type-D}
a \not= 0 \, ,  \qquad  {\cal W}^2 - \frac{b}{a}\, {\cal W} -
\frac{a}{3}\, {\cal G} = 0 \, , \qquad  a \equiv  \Tr {\cal W}^2 \,
, \quad b \equiv \Tr {\cal W}^3 \, ,
\end{equation}
and the differential ones {\em (\ref{charac-U})}, where ${\cal U}$
is the Weyl concomitant:
\begin{equation}
{\cal U} \equiv \frac{{\cal P}({\cal X})}{\sqrt{-{\cal P}^2({\cal
X},{\cal X})}} \, , \qquad  \quad {\cal P} \equiv {\cal W} +
\frac{b}{a} \, {\cal G} \, ,  \label{concomitants-1}\\[2mm]
\end{equation}
${\cal X}$ being an arbitrary SD bivector.
\end{theorem}

A remark on the above theorem. Conditions (\ref{type-D}) state that
the space-time is of Petrov-Bel type D. Then, a previous
result\cite{fsU} implies that, under the first equation in
(\ref{charac-U}) (umbilical condition), the second one (Maxwellian
condition) is a consequence of its real (imaginary) part.

Let us go on the characterization of the ${\cal D}$-metrics in terms
of the sole Ricci tensor. Its traceless part must satisfy the
Rainich conditions:\cite{rai,fs-EM-align} the algebraic ones which
guarantee the type $[(11)(11)]$, and the differential one which
states that the rotation vector determines a closed 1-form. The
Maxwellian character of the structure is then a consequence of the
contracted Bianchi identities and, under the umbilical condition,
the space-time is, necessarily, of Petrov-Bel type D.\cite{fsU}
Moreover the Weyl principal bivector is aligned with the Ricci
tensor. Thus, we obtain the following Rainich-like characterization
of the ${\cal D}$-metrics:
\begin{theorem} \label{Teor-charac-Ricci}
The non vacuum ${\cal D}$-metrics are the space-times whose Ricci
tensor $R$ satisfies:
\begin{eqnarray}  \label{Ricci-alg}
\dif \Tr R = 0, & \qquad \quad   4 \tilde{R}^2 = \Tr \tilde{R}^2\, g
\not= 0 \, , & \qquad \qquad \tilde{R} \equiv R - \frac14 \Tr R \, g \, ,\\
\label{charac-Pi} \sigma[\Pi]=0 \, , & \qquad \qquad \mbox{\rm d}
\Psi[\Pi] = 0 \, ,  & \qquad \qquad \Pi \equiv \frac{1}{\sqrt{\Tr
\tilde{R}^2}} \tilde{R} \, ,
\end{eqnarray}
where $\sigma[\Pi]$ and $\Psi[\Pi]$ are given in {\em
(\ref{sigma})}, and {\em (\ref{PhiPsiPi})}.
\end{theorem}

The original Rainich theory\cite{rai} characterizes the
Einstein-Maxwell solutions (without cosmological constant). In this
case $\tilde{R} = R$ and the first algebraic condition in
(\ref{Ricci-alg}) must be changed to $\Tr R =0$. Moreover, the
algebraic Rainich conditions also impose the energy conditions on
the electromagnetic energy content. This property can be added to
the above theorem with a simple condition: $\tilde{R}(x,x) > 0$,
where $x$ is an arbitrary time-like vector.

It is worth remarking that all the type D vacuum solutions are
${\cal D}$-metrics that are not included in theorem
\ref{Teor-charac-Ricci}. Although theorem \ref{Teor-charac-weyl}
encompasses this vacuum limit, the simplest characterization for the
type D vacuum solutions is $R=0$ and conditions (\ref{type-D}) which
impose an algebraic type D on the Weyl tensor.

\section{Solutions with a Killing tensor: the Kerr-NUT metrics}
\label{sec-Kerr-NUT}

Walker and Penrose\cite{wape} showed that a Killing tensor exists in
the charged Kerr black hole, and Hougston and Sommers\cite{hs1}
proved that, with the exception of the generalized charged
C-metrics, the other ${\cal D}$-metrics have also this property.
From now we call the Kerr-NUT metrics the ${\cal D}$-metrics where a
Killing tensor exists.

Hougston and Sommers\cite{hs2} showed in a subsequent work that the
Killing vector ${\cal Z}$ degenerates (it defines a unique real
Killing vector) if, and only if, the metric admits a Killing tensor.

A later paper by Collinson and Smith\cite{cs} generalized a result
by Floyd\cite{floyd} and Penrose,\cite{penrose} and showed that,
under the assumptions of the papers by Hughston and Sommers, if a
Killing tensor exits, then the space-time also admits a Killing-Yano
2-form. A similar result was also obtained by Stephani.\cite{ste}

In this section we recover all these results, and prove a set of
equivalent conditions that characterize the Kerr-NUT metrics. Next,
we study the Petrov-Bel type D vacuum metrics that admit a
Papapetrou field aligned with the Weyl principal bivector and we
show that they are also the Kerr-NUT space-times.

A {\it Killing-Yano} 2-form is a solution $A_{\alpha \beta}$ to the
equation
\begin{equation} \label{KYd}
\nabla_{(\alpha} A_{\beta) \mu} = 0  \, .
\end{equation}
It is known\cite{kramer} (see also references therein) that the
vector $v = A(t)$ is constant along an affinely parameterized
geodesic with tangent vector $t$. Then, the scalar $v^2$ is a
quadratic first integral of the geodesic equation that,
consequently, defines a second rank {\it Killing tensor}, that is, a
symmetric tensor $K_{\alpha \beta}$ solution to the equation
\begin{equation} \label{Kd}
\nabla_{(\alpha} K_{\beta \mu)} = 0 \, .
\end{equation}
In fact, this Killing tensor $K$ is not but the square of $A$,
$K=A^2$.

Elsewhere\cite{fsY,cfsK} we have given the necessary and sufficient
conditions that a unitary 2-form $U$ must satisfy in order to be the
geometry of either a Killig-Yano tensor\cite{fsY} or a Killing
tensor.\cite{cfsK} A slightly modified version of these results,
more suitable for use here, is the following:
\begin{lemma} \label{kt-ky}
Let $U$ be the geometry of a non null 2-form $A$. Then:

(i) $A$ is a Killing-Yano tensor if, and only if, the bivector
${\cal U}$ satisfies:
\begin{equation} \label{K-Y}
\Sigma =0 \, ,  \quad \quad \dif \chi =0 \, , \quad \quad \dif
\Pi(\chi) = \chi \wedge \Pi(\chi) \, , \quad \quad \xi \wedge \bar
\xi =0 \, ,
\end{equation}
where $\Sigma \equiv \Sigma[U]$ is given in {\rm (\ref{umb})}, $\Pi
= 2 \, {\cal U} \cdot \bar{{\cal U}}$, $\chi = i({\xi}){\cal U}$ and
$\xi = \delta {\cal U}$.

(ii) $K= A^2$ is a Killing tensor if, and only if, $U$ satisfies:
\begin{equation} \label{K-T}
\Sigma =0  \, , \quad \quad \dif \phi =0 \, , \quad \quad \dif
\Pi(\phi) = \phi \wedge \Pi(\phi) \, .
\end{equation}
where $\Phi = \Phi[U]$ is given in {\rm (\ref{Phi-Psi})}.
\end{lemma}
Note that $\Phi$ and $\Pi(\Phi)$ are the real part of $\chi$ and
$\Pi(\chi)$, respectively. Then, it is easy to show that (\ref{K-Y})
implies (\ref{K-T}), in accord with the fact that if $A$ is a
Killing-Yano 2-form then $A^2$ is a Killing tensor.

\subsection{Characterizing the Kerr-NUT space-times}

Let us consider again the family of metrics studied in section
\ref{sec-KV-deltaA}: those admitting a non null CKY 2-form $A$ and
invariant under the complex vector ${\cal Z} = \delta {\cal A}$.

Let us note that ${\cal Z} \wedge \bar{\cal Z}= - 2 \ci Z_1 \wedge
Z_2$ vanishes if, and only if, ${\cal Z}$ determines a unique
Killing direction and, as a consequence of (\ref{ckv}), this fact
equivalently states that $\xi \wedge \bar \xi =0$. From lemma
\ref{kt-ky}, this condition holds when the CKY tensor $A$ is a full
Killing-Yano tensor. We will show now that it is also a sufficient
condition.

Indeed, the two first conditions in (\ref{K-Y}) hold because $A$ is
a CKY tensor. If the last one also holds, $\xi \wedge \bar \xi =0$,
then $A$ will be a full Killing-Yano tensor if its geometry $U$
satisfies the third condition in (\ref{K-Y}).

In order to prove it, we start from $\xi \wedge \bar\xi=0$, which
means that the Killing fields ${\cal Z}$ and $\bar{\cal Z}$ are
collinear and so they differ in a constant $c$, ${\cal Z} = c
\bar{\cal Z}$. Then, if we make the product by ${\cal U}$ and take
into account that, from (\ref{ckv}),  ${\cal Z} = \Omega \xi$, we
obtain:
\begin{equation} \label{chi-1}
\Omega \, \Pi(\chi) = c \,  \bar\Omega \,  \bar\chi \, , \qquad
\Omega \equiv \frac{3}{2} \, e^{- \frac{1}{2} (\phi + \ci \psi)} \,
.
\end{equation}
Note that $\dif \Omega = - \Omega \chi$. Then, the exterior
derivative of the right side in (\ref{chi-1}) vanishes, and the left
side becomes
\begin{equation}
\dif \Omega \wedge \Pi(\chi) + \Omega \dif \Pi(\chi) = 0 \, ,
\end{equation}
and consequently, the third equation in (\ref{K-Y}) holds. Thus, we
have:
\begin{proposition} \label{prop-KN-a}
In the family of non conformally flat metrics admitting a CKY 2-form
$A$ and that are invariant under the complex vector ${\cal Z} =
\delta {\cal A}$, the following three conditions are equivalent:

(i) $A$ is a full Killing-Yano tensor.

(ii) ${\cal Z}$ either vanishes or defines a unique Killing
direction.

(iii) The canonical bivector ${\cal U}$ satisfies $\xi \wedge
\bar\xi=0$, $\xi \equiv \delta  {\cal U}$.\\
In particular, these three conditions are equivalent for the ${\cal
D}$-metrics and characterize the Kerr-NUT space-times.
\end{proposition}

If $A$ is a Killing-Yano 2-form, then $A^2$ is a Killing tensor. Now
we show that, in the family of metrics we are considering and, in
particular, for a ${\cal D}$-metric, the converse is also true.
Thus, let us suppose that the square $K=A^2$ of the CKY tensor is a
Killing tensor and that ${\cal Z}$ is a complex Killing vector.
Then, from the definition of the expansion vector (\ref{Phi-Psi})
and the second and third conditions in (\ref{K-T}) we obtain:
\begin{equation}
\dif i(\delta U)U = \dif i(\delta \!*\!U)\!*\!U = i(\delta U)U
\wedge i(\delta \!*\!U)\!*\!U  \, .
\end{equation}
Thus $i(\delta U)U$ (respectively, $i(\delta \! * \!U)\!*\!U$) is an
integrable direction in the 2-plane $V$ (respectively, $H$). Thus,
lemma \ref{lemma-integ} implies:
\begin{equation}
U(\delta U, \delta \!*\!U) = 0 \, , \qquad  *U(\delta U ,\delta \!
*\!U) = 0 \, ,
\end{equation}
and the definitions (\ref{Phi-Psi}) imply that $\Phi$ and $\Psi$ are
collinear on the principal planes. Then, from proposition
\ref{prop-A-Z} we obtain:
\begin{lemma}
In the family of non conformally flat metrics admitting a CKY 2-form
$A$ and that are invariant under the complex vector ${\cal Z} =
\delta {\cal A}$, if $K=A^2$ is a Killing tensor, then $(\Phi, \xi)
= (\Psi, \xi)=0$.
\end{lemma}

In order to show that $A$ is a Killing-Yano 2-form, we only need to
prove that $\xi \wedge \bar{\xi}=0$ as a consequence of proposition
\ref{prop-KN-a}. We have:
\begin{equation} \label{KT-dif-pi}
\dif \Pi(\xi) = \dif i(\xi) \bar{\cal U} = {\cal L}_{\xi} \bar{\cal
U} - i(\xi) \dif \bar{\cal U} = \chi \wedge \Pi(\chi) - i(\xi) \dif
\bar{\cal U} \, ,
\end{equation}
where we have taking into account that $\bar{\cal U}$ is invariant
under the Killing vector ${\cal Z} = \Omega \xi$ and that $\dif
\Omega = - \Omega \chi$. Now, taking into account the umbilical
equation (\ref{umb}) we can compute
\begin{equation}
i(\xi) \dif \bar{\cal U} = (\xi, \bar{\chi}) \bar{\cal U} -
\bar{\chi} \wedge \Pi(\chi) + \frac32 \, \ci * \! (\xi \wedge
\bar{\xi})  \, .
\end{equation}
Then, substituting this expression in (\ref{KT-dif-pi}) and taking
its real part, we obtain:
\begin{equation}
\dif \Pi(\phi) = \phi \wedge \Pi(\phi) - \frac32 \, \ci * \! (\xi
\wedge \bar{\xi}) \, .
\end{equation}
Finally, from this expression and the third condition in the
characterization (\ref{K-T}) of a Killing tensor, we obtain $\xi
\wedge \bar{\xi}=0$. In real formalism this expression states that
$\delta U \wedge \delta *U = 0$. Thus:
\begin{proposition} \label{prop-KN-b}
Let us consider a non conformally flat metric admitting a CKY 2-form
$A$ such that ${\cal Z} = \delta {\cal A}$ is a complex Killing
vector. Then $K = A^2$ is a Killing tensor, if and only if, the
canonical bivector $U$ satisfies $\delta U \wedge \delta \!*\!U =
0$.
\end{proposition}

From this result and proposition \ref{prop-KN-a} we obtain the
following characterizations of the Kerr-NUT space-times:
\begin{theorem} \label{teo-KN}
The Kerr-NUT space-times are the ${\cal D}$-metrics that satisfy one
of the following equivalent conditions:

(i) The CKY 2-form $A$ is a full Killing-Yano tensor.

(ii) $K=A^2$ is a Killing tensor.

(iii) ${\cal Z}=\delta {\cal A}$ either vanishes or defines a unique
Killing direction.

(iv) The canonical bivector $U$ satisfies $\delta U \wedge \delta \!
*\!U = 0$
\end{theorem}

Note that the last condition in the above theorem involves the sole
Weyl principal bivector $U$. Consequently, if we add this condition
to theorem \ref{Teor-charac-weyl}, we obtain an intrinsic and
explicit characterization of the Kerr-NUT metrics. On the other
hand, this condition may be expressed in terms of the structure
tensor $\Pi$ as $\Phi \wedge \Pi(\Phi) = \Psi \wedge \Pi(\Psi)$,
where $\Phi \equiv \Phi[\Pi]$ and $\Psi \equiv \Psi[\Pi]$ are given
in (\ref{PhiPsiPi}). Then, if we add this condition to theorem
\ref{Teor-charac-Ricci}, we obtain an intrinsic and explicit
characterization of the Kerr-NUT metrics in terms of the sole Ricci
tensor.

\subsection{Type D vacuum solutions with aligned Papapetrou field}

If $\xi$ is a Killing vector, the Papapetrou field $\nabla \xi$ is
closed and, in the vacuum case, it is a solution of the source-free
Maxwell equations.\cite{pap} Metrics admitting an isometry have been
studied by considering the algebraic properties of the associated
Killing 2--form.\cite{faso1,deb1,deb2,mc1,mc2,faso2,stee,lud,faso3}

Elsewhere\cite{fsI1a,fsI1b} we have studied all the Petrov-Bel type
I vacuum space-times admitting a Killing 2--form aligned with a
principal bivector of the Weyl tensor. In the present work we
fulfill a similar study for Petrov-Bel type D metrics. It is known
that in the Kerr geometry this property holds,\cite{faso1,mars} and
we show here that the Kerr-NUT solutions are the type D vacuum
metrics with a time-like Killing 2--form aligned with the Weyl
geometry.

A non null 2-form $F$ has a geometry $U$ if, and only if, its
self-dual part ${\cal F}$ has the direction of ${\cal U}$. This fact
may just be stated as $[F, {\cal U}]=0$.

On the other hand, if $Z$ is a vector field and ${\cal U}$ is a SD
bivector, we have
\begin{equation} \label{uno}
{\cal L}_{Z} {\cal U} = \nabla_{Z} {\cal U} + \frac{1}{2} [\dif Z,
{\cal U} ] + \frac{1}{2} \{ {\cal L}_{Z} g , {\cal U} \} \, .
\end{equation}
From here we obtain the following result:
\begin{lemma} \label{lemma-papa}
A Killing field $Z$ has an associated Papapetrou field
aligned  with the invariant 2-form ${\cal U}$ if, and only if
$\nabla_{Z} {\cal U} =0$.
\end{lemma}

Let ${\cal U}$ be the (invariant) principal bivector of a $\cal
D$-metric. Then, as a consequence of the umbilical equation
(\ref{umb}), the complex Killing vector ${\cal Z} = \Omega \xi$
satisfies:
\begin{equation} \label{nabla-Z-U}
\begin{array}{l}
\nabla_{\cal Z} {\cal U} = \Omega \, \nabla_{\xi} {\cal U}=0  \,
,\cr \nabla_{\bar{\cal Z}} {\cal U} = (\chi,\bar{\cal Z}) {\cal U} +
\Omega \, i({\xi}) i({\bar{\xi}}){\cal G} = \frac{1}{2} \bar{\Omega}
[ \xi \wedge \bar{\xi}- \ci *\!(\xi \wedge \bar{\xi})] \, ,
\end{array}
\end{equation}
where in the last relation we have taking into account that
$(\bar{\chi},{\cal Z})$ vanishes as a consequence of corollary
\ref{corol-D-invariant} and proposition \ref{prop-A-Z}.

Let us consider a real Killing field $Z$ determined by the complex
one ${\cal Z}$, $Z = \mu {\cal Z} + \bar{\mu} \bar{\cal Z}$. From
(\ref{nabla-Z-U}), $\nabla_Z {\cal U}= 0$ if, and only if, $\xi
\wedge \bar{\xi}=0$, that is, $\delta U \wedge \delta\! *\!U = 0$.
Then, from lemma \ref{lemma-papa}, we obtain:
\begin{proposition} \label{prop-papa}
In the set of the $\cal D$-metrics, the necessary and sufficient
condition for a Killing field generated by the complex one ${\cal
Z}$ to have a Papapetrou field aligned with the principal bivector
${\cal U}$ is that $\delta U \wedge \delta \!*\!U = 0$.
\end{proposition}

This proposition and theorem \ref{teo-KN} imply that in the Kerr-NUT
space-times the complex vector ${\cal Z}=\delta {\cal A}$ determines
a unique Killing direction and the associated Killing 2-form is
aligned with the principal bivector. We will see below that in this
case at least one other Killing vector exists and, even the
dimension of the isometry group is bigger than two for the
degenerate family of Kerr-NUT solutions. Nevertheless, for the
generalized C-metrics, only the Killing vectors determined by ${\cal
Z}$ exist. Thus, we have the following result:
\begin{theorem} \label{teor-papa}
The Kerr-NUT solutions are the ${\cal D}$-metrics admitting a
Papapetrou field aligned with the Weyl principal bivector ${\cal U}$
\end{theorem}

We finish by recovering a known result: the ${\cal D}$-metrics
admit, at least, a 2-dimensional commutative group of isometries.
For the generalized C-metrics the complex vector ${\cal Z}$
generates a 2-dimensional commutative algebra. For the Kerr-NUT
metrics it only generates a Killing direction $Z$. But in this case,
if $K$ is the Killing tensor, $Y = K(Z)$ is another Killing vector
that commutes with $Z$.

Indeed, from the Killing tensor equation (\ref{Kd}), for an
arbitrary vector field $Z$ we have:
$${\cal L}_{Z} K - 2 [\nabla Z , K] + {\cal L}_{K(Z)} g =0  \, .$$
The first summand ${\cal L}_{\cal Z} K$ vanishes because $K=A^2$ is
invariant under $Z$. The second summand $[\nabla Z , K]$ also
vanishes because, as a consequence of the proposition
\ref{prop-papa}, the Killing 2-form $\nabla Z$ is aligned with
${\cal U}$, that is, $[\nabla Z , {\cal U}]=0$ and then, $\nabla Z$
commutes with $K = \mu \Pi + \nu g$. Thus, the metric is invariant
under $Y = K(Z)$, a vector field that commutes with $Z$ because $K$
is invariant.

Note that $Z$ and $Y=A^2(Z)$ could define a unique Killing
direction. This means that $Z$ is eigenvector of $A^2$ with constant
eigenvalue. If we impose this condition and consider the expression
of $Z$ and $A$ in terms of $U$, we obtain that these degenerate
Kerr-NUT metrics are characterized by one of the following
conditions:
\begin{equation} \label{KN-degenerate}
i(\delta U)U = i(\delta \!*\!U)U = 0 \, , \qquad  {\rm or} \qquad
i(\delta U)\!*\!U = i(\delta \!*\!U)\!*\!U = 0  \, .
\end{equation}

Under the Kerr-NUT requisite $\delta U \wedge \delta \!*\!U =0$, it
it is easy to show that (\ref{KN-degenerate}) does not hold if
\begin{equation} \label{KN-regular}
i(\delta U)U \wedge i(\delta U)\!*\!U  \not= 0 \, , \qquad  {\rm or}
\qquad i(\delta \!*\!U)U \wedge i(\delta \!*\!U)\!*\!U \not= 0  \, .
\end{equation}

Elsewhere\cite{fsDC} we undertake the intrinsic labeling of the
${\cal D}$-metrics and analyze in detail these degenerate Kerr-NUT
metrics. We show in that work\cite{fsDC} that they admit a
4-dimensional group of isometries with a 2-dimensional commutative
subgroup, accordingly with the known literature on this
subject.\cite{deb-kam-mc,kramer} Taking into account all these
considerations, we can state:
\begin{proposition}
The ${\cal D}$-metrics admit, at least, a commutative 2-dimensional
group of isometries. Let ${\cal U}$ and ${\cal A}$ be the canonical
and the CKY bivectors. Then:

(i) If $\, \delta U \wedge \delta\! *\!U \not=0$ (generalized
C-metrics), the two Killing vectors are determined by ${\cal Z} =
\delta {\cal A}$.

(ii) If $\, \delta U \wedge \delta \!*\!U =0$ and {\rm
(\ref{KN-regular})} (Kerr-NUT regular metrics), one Killing vector
$Z$ is determined by ${\cal Z}$, and the other one is $Y=A^2(Z)$.

(iii) If $\, \delta U \wedge \delta \!*\!U =0$ and {\rm
(\ref{KN-degenerate})} (Kerr-NUT degenerate metrics), the Killing
vector $Z$ determined by ${\cal Z}$ belongs to a 4-dimensional
Killing algebra.
\end{proposition}

\section*{Acknowledgements}
This work has been partially supported by the Spanish Ministerio de
Educaci\'on y Ciencia, MEC-FEDER project FIS2006-06062.

\end{document}